\documentclass[showpacs,prl,twocolumn,floatfix,10pt]{revtex4}

\usepackage{graphicx,float,amsmath,amssymb,psfig}

\begin{document}

\title{Alteration of superconductivity of suspended carbon nanotubes by deposition of organic molecules}

\author{M. Ferrier}
\affiliation{Laboratoire de Physique des Solides, Associ\'e au CNRS,
              Universit\'e Paris-Sud, 91405 Orsay, France.}
\author{A. Yu. Kasumov}
\affiliation{Laboratoire de Physique des Solides, Associ\'e au CNRS,
              Universit\'e Paris-Sud, 91405 Orsay, France.}      
              
\author{V. Agache}        
\affiliation{Institut d'Electronique, de Micro\'electronique et de Nanotechnologiee, UMR CNRS 8520, Cit\'e scientifique, Avenue Poincar\'e, B.P. 69, F-59652 Villeneuve d'Ascq Cedex, France}

\author{L. Buchaillot}
\affiliation{Institut d'Electronique, de Micro\'electronique et de Nanotechnologiee, UMR CNRS 8520, Cit\'e scientifique, Avenue Poincar\'e, B.P. 69, F-59652 Villeneuve d'Ascq Cedex, France}

\author{A-M. Bonnot}
\affiliation{Laboratoire d'Etudes des Propri\'et\'es Electroniques des Solides, CNRS-Grenoble, BP 166, 38042, Grenoble, France}

\author{C. Naud}
\affiliation{Laboratoire d'Etudes des Propri\'et\'es Electroniques des Solides, CNRS-Grenoble, BP 166, 38042, Grenoble, France}
\author{V. Bouchiat}
\affiliation{Centre de Recherches sur les Tr\`es Basses Temp\'eratures, CNRS-Grenoble, BP 166, 38042, Grenoble, France}

\author{R. Deblock}

\affiliation{Laboratoire de Physique des Solides, Associ\'e au CNRS,
              Universit\'e Paris-Sud, 91405 Orsay, France.}
\author{M. Kociak}
\affiliation{Laboratoire de Physique des Solides, Associ\'e au CNRS,
              Universit\'e Paris-Sud, 91405 Orsay, France.}           
\author{M. Kobylko}
\affiliation{Laboratoire de Physique des Solides, Associ\'e au CNRS,
              Universit\'e Paris-Sud, 91405 Orsay, France.}   
\author{S. Gu\'eron}
\affiliation{Laboratoire de Physique des Solides, Associ\'e au CNRS,
              Universit\'e Paris-Sud, 91405 Orsay, France.}            
\author{H. Bouchiat}
\affiliation{Laboratoire de Physique des Solides, Associ\'e au CNRS,
              Universit\'e Paris-Sud, 91405 Orsay, France.}

\date{\today}

\begin{abstract}
We have altered the superconductivity of a suspended rope of single walled carbon nanotubes, by coating it with organic polymers.
Upon coating, the normal state resistance of the rope changes by less than 20 percent. But superconductivity, which on the bare rope shows up as a substantial resistance decrease below 300 mK, is gradualy suppressed. We correlate this to  the suppression of radial breathing modes, measured with Raman Spectroscopy on suspended Single and Double-walled carbon nanotubes. This points to the breathing phonon modes as being responsible for superconductivity in carbon nanotubes.
\end{abstract}


\maketitle

Carbon nanotubes have been heralded as model systems to explore one dimensionnal (1D) conductors, in which electron-electron interactions lead to a non conventional ground state, the Luttinger liquid \cite {Egger}. In particular, the single particle density of states is depressed at low energy, with a power law whose exponent depends on the interaction strength. The experimental observation of the power law suppression of the tunnel conductance at low energy was therefore interpreted as a proof of this ground state and of the strength of repulsive electron electron interactions in Single Walled Carbon Nanotubes (SWNT). The discovery of superconductivity in suspended individual ropes of SWNT therefore came as a big surprise \cite{Kociak2001,Kasumovreview}. As for the case of metallofullerene molecules \cite{Dresselhaus}, the question of which phonon modes give rise to the attractive interaction leading to superconductivity is still unsettled. Whereas S\'ed\'eki {\it et al.} \cite{Sédéki} and Gonzalez \cite{Gonzalez2001,Gonzalez2002} consider the coupling to optical phonon modes, De Martino and Egger \cite{Egger1,Egger2} conjecture that attractive interactions can be mediated by low energy phonon modes,  and specifically the radial breathing modes (RBM), which are the compression and dilatation modes of carbon nanotubes. They find that the attractive interaction may be strong enough to overcome the repulsive interactions in SWNT, especially in ropes of SWNT where the Coulomb interaction can be screened because the SWNT are packed so closely together. Their theory \cite{Egger2} also could reproduce the temperature dependence of the superconducting transition observed in ropes.

To experimentally test the role played by the phonon modes on superconductivity, we have gradually coated a suspended rope of SWNT with organic material. We show that superconductivity is gradually destroyed. Parallel Raman experiments show that the radial breathing modes are affected by coating, thereby hinting to these modes as playing a major role in the superconductivity of carbon nanotubes. The experiments described hereafter also help define the criteria required to observe a superconducting transition in carbon naotubes.

\begin{figure}
\includegraphics[clip=true,width=7.5cm]{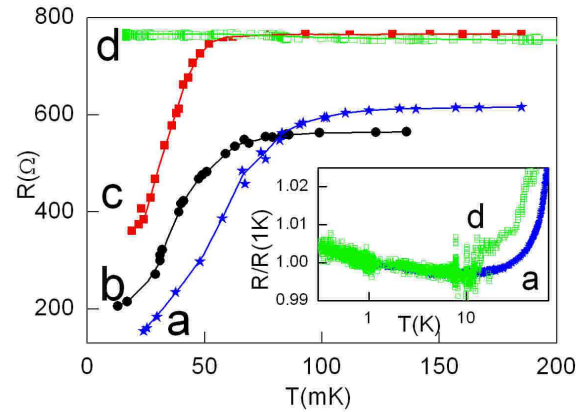}
\caption{Temperature dependence of the resistance below 400 mK of the 10 nm wide, 1~$\mu$m-long rope containing 40 SWNT. a) Before coating. b) After benzene-dithiol deposition. c) After additional coating with PMMA. d) After completely covering the rope and the slit with PMMA. Lines are guides to the eyes. Curve {\it d} is the recorded differential resistance at zero current as the sample is heated. Curves {\it a, b} and {\it c} are constructed by extracting the differential conductance at zero current from a $dV/dI$ Vs $I$ curve, at different temperatures. This ensures a stable temperature of the sample, which, for reasons which may have to do with the low heat conduction of small superconducting samples, was slow in reaching thermal equilibrium.
Inset: Comparison at higher temperature. The upturn in resistance around 10 K is the signature of twiston modes characteristic of well ordered carbon nanotube ropes. Both curves, before and after complete coating are similar, except for the unstable resistance in stage {\it d} above 10 K.}
\label{RdeT}
\end{figure}

The sample whose superconductivity we have altered is a rope containing roughly 40 SWNT, as determined from its diameter (10 nm) measured in a Transmission Electron Microscope. This rope is produced by the arc discharge method followed by purification \cite {Vaccarini, Journet}. The rope is suspended across a $1~\mu m$ wide slit etched in a suspended silicon nitride membrane, and well connected to non superconducting (normal) electrodes {\it via} the nanosoldering technique \cite{Kasumovreview}. The electrodes are a trilayer of 5 nm $Al_20_3$, 3 nm Pt, and 200 nm Au deposited in this order on the membrane. The superconducting transition of the uncoated rope was reported in \cite {Kasumovreview}, and is plotted as curve (a) in Fig. \ref{RdeT}. The rope's resistance decreases from $600~\Omega $ above 120 mK to $150~\Omega$ at 25 mK. This transition is weakened in a magnetic field, and suppressed above a field of 0.6~T.

The sample was then modified in successive steps. We first (stage {\it b}) coated it at room temperature with a drop of benzenedithiol. In step {\it c}, we coated the rope with Poly(methyl methacrylate) (PMMA). In step {\it d} even more PMMA was added, so that the entire slit was covered with PMMA.
The trend upon successive coatings, presented in Fig. \ref{RdeT}, is to have a modified normal state (high temperature) resistance, which varies between 550 and 750 $ \Omega$, and a superconducting transition which weakens as the nanotube rope is coated. The transition temperature T*, defined as the temperature at which the resistance starts to decrease, which would be the transition temperature of a (hypothetical) corresponding 3D superconductor, shifts from 120 mK to 60 mK in curves {\it a} to {\it c}. In curve {\it d}, the coating is so important that no transition is visible down to less than 15 mK. The resistance even increases by $0.5~\%$ between 200 mK and 16 mK.
  
As discussed in previous papers (\cite {Kociak2001,Kasumovreview}), the lowest resistance state is not a zero resistance state for two reasons. First, a contact or Landauer resistance of at least $R_Q= h/(2e^{2})=12.9~ k\Omega$ always exists between a conduction channel and a normal reservoir, so that for this rope of N=40 nanotubes each containing two conduction channels, the smallest contact resistance in the superconducting state would be $R_Q/(2N)=160~ \Omega$, and not zero (see \cite{Klap} for a recent experimental confirmation of the contact resistance in a NSN structure). Second, quantum phase slips should also be present at low temperature, contributing an additional resistance \cite{Giordano}. 

\begin{figure}
\includegraphics[clip=true,width=8 cm]{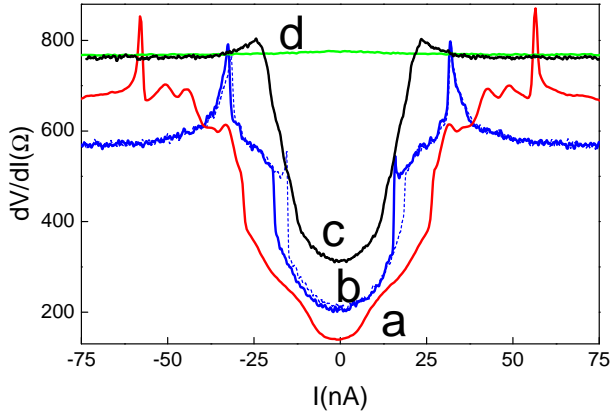}

\caption{Differential resistance of the nanotube rope at stages {\it a} through d. Curve a was taken at 28 mK with an ac excitation of 1 nA, curve {\it b} at 14 mK with i$_{\texttt{ac}}$=0.1 nA, curve {\it c} at 15 mK with i$_{\texttt{ac}}$=0.1 nA, and curve {\it d} at 15 mK with i$_{\texttt{ac}}$=1 nA. Both the up (dashed) and down current sweep directions are plotted for curve {\it b}, illustrating the asymmetry of these current-biased curves.}
\label{dVdI}
\end{figure}

The gradual weakening of the rope's superconductivity is even more clearly seen in the differential resistance curves plotted in Fig. \ref{dVdI}. The differential resistance of the uncoated rope (curve {\it a}) is typical of a 1D superconductor \cite{Bezryadin}(a wire in which the superconducting coherence length is larger than the diameter): the differential resistance is smallest at zero current bias, and increases  with jumps and peaks typical of phase slips as the current through the rope is increased. At high current bias, the normal state resistance is recovered. We define a critical current $I_c$ as the current above which the differential resistance jumps from a low value to a high value (in analogy to the bulk critical current, at which the differential resistance jumps from zero to finite resistance). We also define $I_{c}$$^*$ as the current of the last resistance peak, above which all traces of superconductivity disappear. As shown on the example of curve {\it b}, the dV/dI(I) curves are hysteretic, and we take the largest values to be the switching values. They are given in the table for successive coatings, and are seen to decrease as the rope is coated with an increasing amount of organic material.

\begin{figure}
\includegraphics[clip=true,width=8 cm]{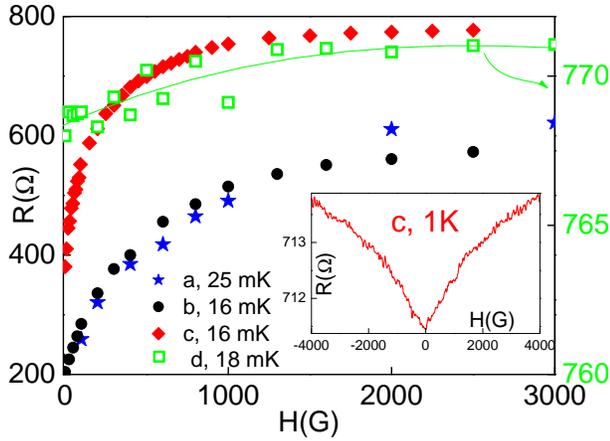}
\caption{Magnetoresistance of the sample at the different coating stages. A large positive magnetoresistance is seen at low temperature for stages {\it a,b and c}. At stage {\it d} the magnetoresistance is still positive, but small (0.1 \%), identical to the magnetoresistance of stage {\it c} at 1 K. Left resistance scale: Samples a, b, and c at low temperature. Right resistance scale is relative to sample d that shows a very small positive magnetoresistance. The line is a guide to the eye.}
\label{RdeH}
\end{figure}

The resistance variations with magnetic field are shown in Fig. \ref {RdeH}. No qualitative difference is noticeable between curves {\it a}, {\it  b} and {\it c}, which show a large (more than $50~ \%$) positive magnetoresistance (the resistance increases as the magnetic field destroys superconductivity). The amplitude of this magnetoresistance decreases as the temperature is increased, but persists up to 1 K, which is nearly 10 $T^{*}$. The critical field, above which the differential resistances do not show a dip at low current, is 0.6 T. Interestingly, a small ($0.5 ~\%$) positive magnetoresistance is visible at stage {\it d}, where no other trace of superconductivity is measureable,  and has an amplitude similar to the magnetoresistance of stage {\it b} at 1 K, as shown in the inset of Fig \ref {RdeH}. Superconducting fluctuations over such broad temperature ranges are typical of low dimensional superconductivity.

\begin{table}[bp] 

\begin{tabular}{|c|c|c|c|c|}
\hline
 &  $R_{1 K}$ & T* & $I_c$ & $I_{c}$$^*$ \\
\hline
$R4a_{PtAu}$& 627 $\Omega$&120 mK&28 nA&58 nA \\
 \hline
$R4b_{PtAu}$& 564 $\Omega$&80 mK& 19 ~nA& 32~nA\\ \hline
$R4c_{PtAu}$& 763 $\Omega$&60 mK&*** &24 ~nA\\ \hline
$R4d_{PtAu}$& 767 $\Omega$&$<$ 15 mK&0 nA&0 nA\\ \hline
\end{tabular}

\caption{Rope characteristics at different modification stages. The rope contains roughly 40 nanotubes, is  $1~ \mu m$ long.
{\it a}: before coating, as reported in ref \cite {Kasumovreview}; {\it b}: After coating with benzenedithiol; {\it c}: after further coating with PMMA; {\it d}: Additional PMMA. $I_c$ and $I_{c}$$^*$ are the currents at which the first and last differential resistance jumps occur.}
\label{modiftable}
\end{table}

\begin{figure}
\includegraphics[clip=true,width=7 cm]{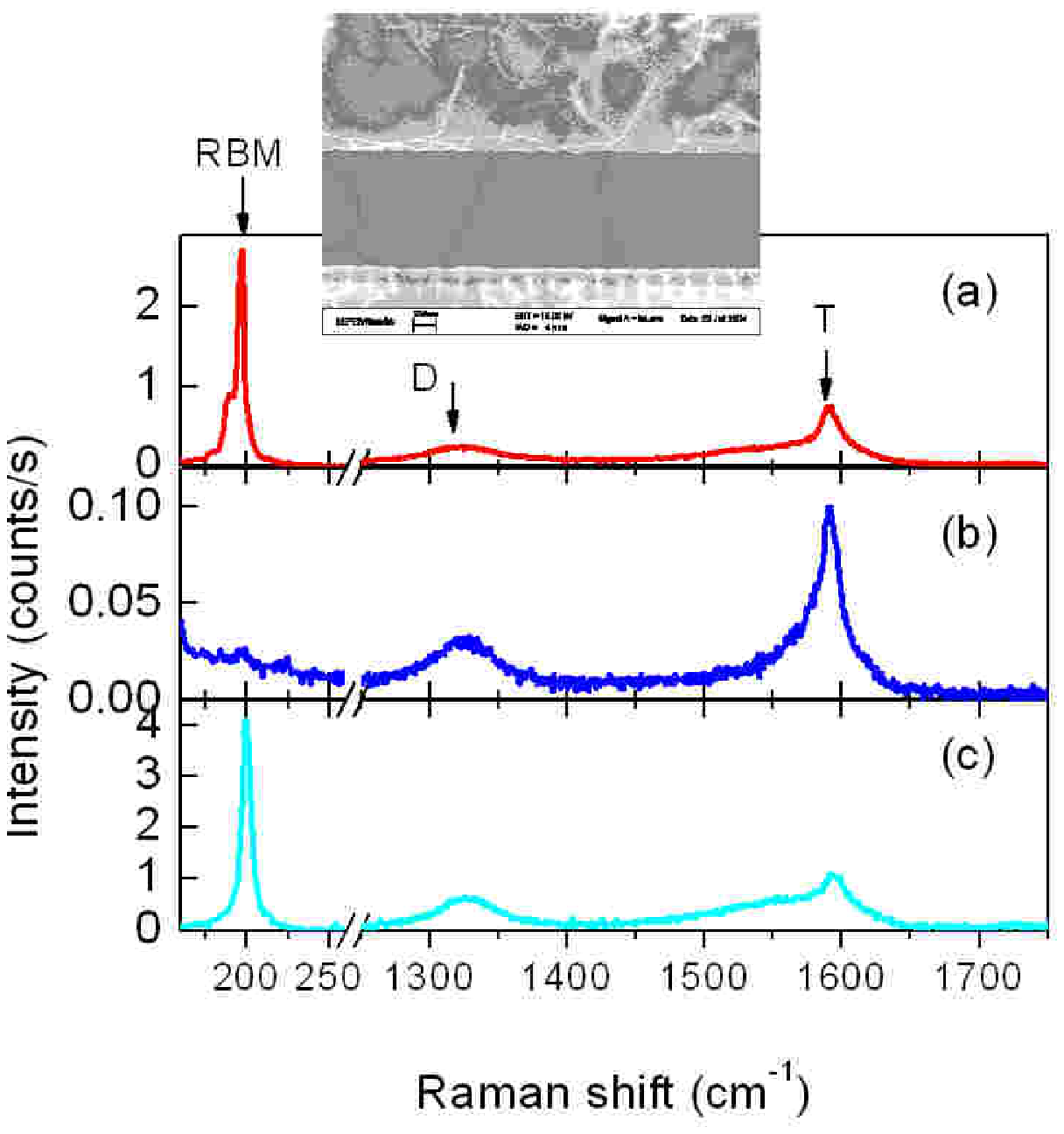}
\caption{
Influence of PMMA on the Raman spectra of suspended SWNTs. Raman spectra obtained at room temperature with a He-Ne laser radiation (1.96 eV). The laser probe area is roughly $1 ~ \mu m^2$. From top to bottom: (a) Raman spectra before deposition, (b) after deposition, and (c) after removal of PMMA in acetone. $RBM$, $D$, and $T$ respectively stand for Radial Breathing Mode, disordered graphite mode, tangential modes. Inset: SEM micrograph of a typical sample of CVD-grown tubes.}

\label{Raman}
\end{figure}

We now show how the disappearance of superconductivity can be correlated to the blocking of the RBM of suspended carbon nanotubes upon coating with organic material, as investigated with Raman spectroscopy. 
The nanotubes grown for this purpose were synthesized by chemical vapor deposition (CVD) in methane and hydrogen, with the assistance of a hot (2000~$^\circ$~C) tungsten filament placed 1 cm away from the substrate \cite {Nanolett}, which also was a suspended silicon nitride membrane with a slit. A thin Co layer served as the catalyst. The substrate temperature was about 800 $^\circ$~C during synthesis. Fig. \ref{Raman} shows a Field Emission Scanning Electron Micrograph of carbon nanotubes grown in this manner. It is typical that thinner ropes grow straighter than thicker ropes. The Transmission electron microscopy and Raman spectroscopy we have performed show that the nanotubes are  single- and double- walled carbon nanotubes (mostly SWNT), sometimes assembled in small ropes. Resonant Raman measurements are performed using the 1.96 eV radiation of a 633 nm He-Ne laser  with a $1~\mu m$$^2$ spot size, so that only a few suspended tubes are in resonance with the laser. Typical Raman spectra contain RBM in the 100-260 cm$^{-1}$ wavelength range (12.5 to 32.4 meV, which, given the theoretical law of 27.8/d meV \cite{diameter}, where d is the SWNT diameter in nm, corresponds to diameters between  0.9 and 2.2 nm), and tangential modes around 1560-1600 cm$^-1$. The low Raman signal intensity, along with the observation of one or two RBM point to the presence of small ropes or single isolated tubes \cite{Saito}.
We attribute the strong intensity of the RBM (relative to the intensity of the tangential modes) to the fact that the tubes are suspended, and therefore do not interact through Van der Waals interaction with the substrate \cite{Sauvajol}.  Also, the low intensity of the band at 1320 cm$^-1$ which characterizes structural disorder suggests that these tubes are relatively defect free.
In Fig. \ref{Raman}, we follow the evolution of the phonon modes with coating by comparing Raman spectra which have a same RBM frequency, before (a) and after (b) PMMA deposition, and after removal of PMMA in acetone (c). Although these spectra do not correspond to the very same tubes, because of the absence of precise sample repositioning, the tubes are nonetheless of roughly the same diameter, since we the RBM have the same frequency. The 218 cm$^-1$ frequency of Fig. \ref {Raman} is typical of Raman resonant RBM mode of metallic tubes \cite{Kataura}. The broad and asymmetric tangential band around 1530 cm$^-1$ is the signature of the Breit Wigner Fano electron-phonon intertube interaction specific to metallic nanotubes \cite{Jiang}, while the sharper peak at 1590 cm$^-1$ is typical of tangential modes.
With coating an overall decrease of the Raman intensity is observed, and the RBM intensity is much smaller than the intensity of the tangential mode.
After removal of the PMMA with acetone, the intensity of the metallic RBM is clearly recovered, as is the broadened tangential mode. The spectra and ratios shown in the figure are typical of what we find in many measurements on this sample and others.

This experiment shows that coating tubes with PMMA suppresses its RBM. The fact that the RBM Raman signal is much larger for suspended tubes than for tubes on a substrate, has also been observed in recent Scanning Tunneling Microscopy experiments on suspended nanotubes \cite{LeRoy}. The authors found spectroscopic signature of the RBM modes only on the suspended portions of the tubes, and not on the contact regions.

In the transport experiment on the rope containing 40 tubes, it is therefore likely that the RBM of the tubes on the surface of the rope are suppressed because they are coated by PMMA. There are 20 such outer tubes. The other 20 tubes are inside the rope and are therefore not coated by PMMA. It is however likely that the RBM of those inner tubes should also be suppressed at low temperature, for a different reason: the large ($\epsilon=1~\%$) thermal contraction of PMMA between room temperature and low temperature imposes a large pressure on the nanotube rope. Given a Young modulous E of $3\times10^{9}N/m^{2}$ in plexiglass, the tubes within the rope experience a pressure larger than $P=E \epsilon D/d$, with D the diameter of the PMMA coating layer and $d=10~ $ nm the rope diameter. For a capping layer roughly $1~\mu m$ thick, we estimate $P~\approx $3~GPa. Merlen {\it et al.}  have found that the Raman intensity of the RBM of SWNT under isostatic pressure at room temperature is weakened above 1 GPa, divided by two at 10 GPa, and by ten at 12 GPa \cite{Merlen}, so that it is reasonable to assume that the RBM of the inner tubes are suppressed as well at low temperature.

Interestingly, although the RBM are clearly suppressed in the coated rope, it seems as though the twiston modes are not affected. Twistons, the long-wavelength torsional modes of nanotubes, are known to backscatter electrons efficiently, as demonstrated by a measurable linear temperature dependence down to 100-10 K, below which  a resistance increase takes place \cite{twistons}. As shown in Fig. \ref{RdeT}, before coating and after complete coating, the resistance decrease and upturn with decreasing temperature seem similar, suggesting that the twiston modes are not modified by coating. A qualitative picture is thus that the micron thick PMMA layer around the ropes induces a strong radial pressure at low temperature but barely a longitudinal pressure. Therefore the  longitudinal modes (stretching modes) are not affected by coating whereas the radial breathing modes are. Twistons, a combination of both, are affected to a lesser extent. More generally, it is expected that optical modes should be much less affected by coating than acoustic modes. 
This experiment therefore seems to confirm the theoretical picture put forth in \cite{Egger1} that the RBM more than the longitudinal modes cause superconductivity. As conjectured, the twiston modes backscatter electrons efficiently but do not participate in the phonon-induced forward scattering of electrons responsible for the attractive interaction.

We argue that the progressive destruction of superconductivity is not due to an increase of disorder in the sample, for two reasons. The first is the small relative resistance change (less than 1~$\%$) between stage {\it c} and {\it d}, which shows that disorder is not qualitatively different. The destruction of superconductivity in stage {\it d} is therefore caused by some other mechanism.
 For the same reason, the destruction of superconductivity cannot be attributed to a pressure-induced modification of the nanotubes' band structures, turning metallic tubes into semi-conducting ones, for this would change by a large amount the sample resistance. The second fact pointing to small disorder in the rope is the very low shot noise that we have measured in this rope at 1 K at stage {\it c}: The shot noise power was $S_I=2eI/120$, much less than the shot noise power $S_I=2eI/3$ expected for a coherent diffusive sample \cite{Meydi}. Such a large reduction of the shot noise power could be the sign that transport through the rope is almost ballistic \cite {Roche}.

We have shown that the suspended character of nanotubes is a crucial condition to observe superconductivity. Coating the nanotubes suppresses the radial breathing modes,  suggesting that these phonon modes play a role in the superconductivity of carbon nanotubes. The experiments conducted down to low temperature to this day outline the stringent requirements to observe superconductivity: Nanotubes should not be coated by polymers, and not in contact with a substrate. They should be long enough in order to avoid the inverse proximity effect, which causes the proximity to normal contacts to destroy the superconductivity in the rope. Finally, there should be efficient screening of the repulsive interaction in nanotubes, so that superconductivity should develop most in tubes that are assembled in a rope.


\newpage

\pagebreak

\end{document}